\documentstyle[aps,prb,twocolumn]{revtex}
\draft
\begin{document}
\title{Ballistic transport through chaotic cavities:
Can parametric correlations and the
weak localization peak be described by a
Brownian motion model?}
\author{Jochen Rau}
\address{Max-Planck-Institut f\"ur Kernphysik,
Postfach 103980, 69029 Heidelberg, Germany}
\date{Submitted to Phys. Rev. B, August 23, 1994}
\maketitle
\begin{abstract}
A Brownian motion model is devised on the manifold of
$S$-matrices, and applied to the calculation of
conductance-conductance correlations and of the
weak localization peak.
The model predicts that (i) the correlation function in $B$ has the
same shape and width as the weak localization peak;
(ii) the functions behave as $\propto 1-{\cal O}(B^2)$,
thus excluding a linear line shape; and
(iii) their width increases as
the square root of the number of channels in the leads.
Some of these predictions agree with experiment and with other
calculations only in the limit of small $B$ and
a large number of channels.
\end{abstract}
\pacs{72.10.Bg, 72.20.My, 05.45.+b, 05.40.+j}
\medskip
\narrowtext
\section{INTRODUCTION}
The experimental observation of quantum interference effects
in submicron devices has triggered rapidly growing interest
in the properties of mesoscopic systems.
\cite{reviews}
Many investigations have focused on electronic transport through such
microstructures and, in particular, on the behavior of the conductance $G(X)$
as a function of some parameter $X$.
(Two examples for $X$ are the Fermi wave vector $k$
or an external, time-independent magnetic field $B$.)
Both in
diffusive transport through quasi-one-dimensional disordered structures
\cite{reviews}
and (more recently) in ballistic transport through chaotic cavities
\cite{marcus} one has observed the following two phenomena.
(i) As the parameter is being varied, the conductance
fluctuates irregularly around its mean value.
These fluctuations have a typical magnitude
of the order $e^2/h$.
This magnitude is independent
of material, geometry or degree of disorder of the probe
(`universal conductance fluctuations');
however, it does change depending
upon the presence or absence of time-reversal symmetry.
(ii) The average conductance at zero magnetic field is
consistently lower than the average conductance at large magnetic
field, the difference being again of the order $e^2/h$
(`weak localization').
These two phenomena are well accounted for not only
by microscopic theories,\cite{micro}
but also by macroscopic random matrix theories.
\cite{transfer,iwz,smatrix}
The success of the latter clearly demonstrates
the universal nature of the observed effects.

The universality of the {\em magnitude} of conductance
fluctuations and of the weak localization correction being established,
the obvious question arises whether also the
{\em frequency spectrum} (`power spectrum')
of the conductance fluctuations, as well as
the {\em width} and {\em shape}
of the weak localization peak, are universal.
The power spectrum is the Fourier transform of the
conductance-conductance correlation function
\begin{equation}
\label{corr_def}
C(\Delta X):=\langle \delta G(X) \delta G(X+\Delta X) \rangle
\quad,
\end{equation}
where $\delta G:=G-\langle G\rangle$ denotes the
deviation of the conductance from its mean;
while width and shape of the weak localization peak are
encoded in the function
\begin{equation}
\label{weak_def}
W(B):=\langle G\rangle (B) - \langle G\rangle(B=\infty)
\quad.
\end{equation}
(The averages are taken over an ensemble of probes
with -- in the case of disordered structures --
different impurity configurations,
or -- in the case of chaotic cavities --
different shapes.
By a widely accepted ergodicity hypothesis, ensemble averages
are equivalent to the average over
some parameter such as $k$ or $B$.)
Both $C(0)$ and $W(0)$ are well known;
\cite{micro,transfer,iwz,smatrix}
the objects to be investigated are the ratios
$C(\Delta X)/C(0)$ and $W(B)/W(0)$.

In this paper I focus on the ballistic regime; moreover,
I assume the ballistic cavities to be classically chaotic.
For such systems the ratios
$C(\Delta X)/C(0)$ and $W(B)/W(0)$ have been calculated
semiclassically.
For example, semiclassical analysis
yields a Lorentzian,
\begin{equation}
C(\Delta k)/C(0)=[1+(\Delta k/\gamma_{\rm cl})^2]^{-1}
\quad,
\label{mom_corr}
\end{equation}
or a Lorentzian-squared,
\begin{equation}
C(\Delta B)/C(0)=[1+(\Delta B/\alpha_{\rm cl}\phi_0)^2]^{-2}
\quad,
\label{mag_corr}
\end{equation}
respectively, for the correlation function.\cite{jalabert_corr}
(Here $\gamma_{\rm cl}$ and $\alpha_{\rm cl}$ are characteristic
inverse length and characteristic inverse enclosed area,
respectively, of the classical trajectories;
$\phi_0=hc/e$ is the elementary flux quantum.)
The semiclassical result for
$C(\Delta B)/C(0)$ has been checked experimentally and appears to
be in excellent agreement with the data.\cite{marcus}
For the weak localization peak,
semiclassical theory predicts a Lorentzian,
\cite{baranger_weak}
\begin{equation}
W(B)/W(0)=[1+2(B/\alpha_{\rm cl}\phi_0)^2]^{-1}
\quad.
\label{weak_peak}
\end{equation}
This prediction, too, agrees well with experiments and with
numerical simulations.\cite{marcus,chang}
We observe that the correlation function in $B$ and the weak
localization peak look almost identical.
Indeed, for small $B$
\begin{equation}
C(B)/C(0) \approx W(B)/W(0) \approx 1-2(B/\alpha_{\rm cl}\phi_0)^2
\quad;
\label{approx_equal}
\end{equation}
only the tails at large $B$ differ.

The above semiclassical calculations involve various approximations.
In addition to (i) the semiclassical limit proper, it is assumed
that (ii) the magnetic field is sufficiently weak so as to affect
only phases, but not the classical trajectories, and that
(iii) in the case of weak localization, only symmetry-related
paths contribute (`diagonal approximation').
The first two assumptions are expected to be justified only in the
limit of weak fields and a large number of channels;
while the third assumption has in fact been shown to be too crude.
\cite{baranger_weak}
Moreover, a recent study \cite{hastings} has revealed that important
contributions to weak localization stem from diagrams which appear to
have no semiclassical analogues.
In view of these theoretical limitations it is quite remarkable
that semiclassics is so successful in describing the shapes of $C$ and $W$.
Nevertheless,
it seems highly desirable to derive parametric
correlation functions and the weak localization peak by
some alternative method.

The above results (\ref{mom_corr}), (\ref{mag_corr}) and
(\ref{weak_peak}) are universal in the sense that the
geometry of the probe enters only via the characteristic scale
$\gamma_{\rm cl}$ or $\alpha_{\rm cl}$.
Such universality suggests that the phenomena
in question
should lend themselves to a treatment in the framework of
random matrix theory.
Indeed, Pluhar et al. \cite{pluhar} showed that the
Lorentzian shape of the weak localization peak can be derived
with Hamiltonian random matrix theory.
In addition, they succeeded in proving that the width of the
weak localization peak
increases essentially as the square root of the number of channels.
(This is a result which in a semiclassical theory can be inferred
only indirectly.)
As regards parametric correlations,
Altland \cite{altland}
employed Hamiltonian random matrix theory to show
that in the diffusive regime and quasi-one-dimensional limit,
$C(\Delta E)$ is a Lorentzian. ($E$ denotes the Fermi energy of the
electrons.)
Other macroscopic calculations of parametric correlations have
so far focused on
correlations of the level density, rather than on correlations
of the conductance.\cite{simons}
Their success certainly encourages further applications
of macroscopic random matrix models.

A particularly simple and (arguably) generic model for any
kind of parametric dependence is furnished by
Dyson's Brownian motion model.\cite{dyson}
It is based on the assumption that
variation of the parameter $X$ amounts to a random walk on
the manifold of Hamiltonians, a process
governed by a diffusion
(or Fokker-Planck) equation.
Beenakker \cite{beenakker} successfully employed this model
to calculate density-density correlations, and
suggested that it might also be applied to the response of
transmission eigenvalues (and hence of the conductance)
to an external perturbation.
This is difficult, however,
because the conductance -- in contrast to the level density --
cannot be easily expressed as a
function on the manifold of Hamiltonians.
This difficulty can be avoided if one considers random $S$-matrices
rather than random Hamiltonians.
Random $S$-matrix theory,\cite{smatrix}
while consistent with Hamiltonian calculations \cite{iwz}
in the limit of a large number of channels,
allows a much more direct calculation of conductance properties.
Therefore, I propose that the Brownian motion model be
reincarnated
as a random walk on the manifold of $S$-matrices.\cite{equivalence}
To devise such a Brownian motion model for $S$-matrices, and to study
its implications for conductance-conductance correlations
and the weak localization peak,
is the purpose of this paper.
\section{THEORETICAL PRELIMINARIES}
I will first review some basic definitions.
Electronic transport through a microstructure constitutes a
quantum mechanical scattering problem.\cite{landauer}
Scattering at a probe with two leads, each with $N$ channels,
is described by the $S$-matrix
\begin{equation}
S=\left(\begin{array}{cc}
  r&t'\\ t&r'
  \end{array}\right)
\end{equation}
where $t,t'$ and $r,r'$ are the $N\times N$ transmission and
reflection matrices, respectively.
Both $(t^\dagger t)$ and $(t'^\dagger t')$ have the same set of
eigenvalues $\{\tau_a\in[0,1]\}$. In terms of these eigenvalues, the
$S$-matrix may be parametrized
\begin{equation}
S=\left( \begin{array}{cc}
  v^{(1)} & 0 \\
  0 & v^{(2)}
   \end{array} \right)
  \left( \begin{array}{cc}
  -\sqrt{1-\tau} & \sqrt\tau \\
  \sqrt\tau & \sqrt{1-\tau}
   \end{array} \right)
 \left( \begin{array}{cc}
  v^{(3)} & 0 \\
  0 & v^{(4)}
   \end{array} \right)
\quad
\end{equation}
with $\tau=\mbox{diag}(\tau_a)$
and the $\{v^{(i)}\}$ unitary $N\times N$ matrices.
\cite{param}
The $\{\tau_a\}$ and $\{v^{(i)}\}$ will be referred to as
`radial' and `angular' degrees of freedom, respectively.
In the absence of time-reversal symmetry (`unitary case,' $\beta=2$)
the $\{v^{(i)}\}$ are
arbitrary; whereas the presence of time-reversal symmetry
(`orthogonal case,' $\beta=1$) imposes the
additional constraint $S^T=S$ and hence
$v^{(3)}=v^{(1)T}$, $v^{(4)}=v^{(2)T}$.\cite{symplectic}
Depending upon the symmetry,
the manifold of $S$-matrices has the dimension
\begin{equation}
d_\beta:= \left\{
 \begin{array}{ll}
  2N^2+N & \,:\, \beta=1 \\
  4N^2 & \,:\, \beta=2
  \end{array} \right.
\quad.
\end{equation}
The manifold is endowed with a metric
\begin{equation}
g({\rm d}S_1, {\rm d}S_2) := \mbox{tr} ({\rm d}S_1^\dagger {\rm d}S_2)
\end{equation}
which is invariant under the group action
${\rm d}S_i\to U{\rm d}S_i V$ ($U,V$ unitary).
With respect to this metric,
radial variations ($\tau_a\to\tau_a+{\rm d}\tau_a$) and
angular variations ($v^{(i)}\to v^{(i)}+{\rm d}v^{(i)}$) of
the $S$-matrix are orthogonal;
the metric tensor $(g_{\mu\nu})$ is thus built of two blocks,
one `radial' and one `angular.'
The radial part of the metric tensor reads
\begin{eqnarray}
g_{ab} &:=& g(\partial S/\partial\tau_a, \partial S/\partial\tau_b)
\nonumber \\
&=& {1\over 2\tau_a(1-\tau_a)} \delta_{ab}
\quad,
\end{eqnarray}
with inverse
\begin{equation}
g^{ab}=2\tau_a(1-\tau_a) \delta^{ab}
\quad.
\label{radial_metric}
\end{equation}
The metric defines an invariant volume element
(Haar measure)
\begin{eqnarray}
{\rm d}\mu(S) &=&
\sqrt{|\det g|} \prod_\alpha {\rm d}x^\mu
\nonumber \\
&=& J_\beta(\{\tau\}) \prod_a {\rm d}\tau_a\cdot \prod {\rm d}[\mbox{angles}]
\quad
\label{volume}
\end{eqnarray}
where $J_\beta$ is the Jacobian \cite{smatrix}
\begin{equation}
J_\beta =
\left\{ \begin{array}{ll}
  \prod_{a<b}|\tau_a-\tau_b|\cdot\prod_c 1/\sqrt{\tau_c} &:\,\beta=1 \\
  \prod_{a<b}|\tau_a-\tau_b|^2                           &:\,\beta=2
  \end{array}
\right.
\quad.
\end{equation}
(The exact form of the angular factor need not concern us.)
This volume element in turn
permits the definition of probability densities $P(S)$,
with normalization
\begin{equation}
\int{\rm d}\mu(S)\,P(S)=1
\quad.
\end{equation}

In the above parametrization $\{\tau_a,v^{(i)}\}$,
the conductance can be easily expressed
as a function of $S$.
According to Landauer's formula
\cite{landauer}
it reads
\begin{equation}
G=(2e^2/h)\sum_a \tau_a
\end{equation}
(with the factor $2$ accounting for spin).
The sum $T:=\sum_a\tau_a$ is
called the dimensionless conductance.
Assuming a uniform probability density on the manifold
of $S$-matrices,\cite{coupling} $P(S)=const$, one obtains the
well-known results \cite{smatrix}
\begin{equation}
\langle T\rangle_\beta ={\beta N^2\over d_\beta} N
= {N\over2} - \delta_{\beta 1} {N\over 4N+2}
\end{equation}
and
\begin{equation}
\mbox{var}_\beta (T)=\left\{
 \begin{array}{ll}
  {N(N+1)^2\over (2N+1)^2 (2N+3)} & \,:\, \beta=1 \\
  {N^2\over 4(4N^2-1)} & \,:\, \beta=2
 \end{array}
 \right.
\quad.
\end{equation}
For $N\to\infty$, these imply the universal magnitude of weak localization,
$\langle T\rangle_{1} - \langle T\rangle_{2} \to (-1/4)$,
and of conductance
fluctuations, $\mbox{var}_\beta (T)\to (1/8\beta)$.
\section{BROWNIAN MOTION MODEL}
In the Brownian motion model
the probability density is no longer
uniform and
acquires an explicit dependence on the parameter $X$.
As $X$ is being varied, $X\to X+\Delta X$,
the system executes a random walk on
the manifold of $S$-matrices and,
consequently, the probability density spreads according to
a diffusion equation.
In order to find the form of this diffusion equation,
let us consider -- as a simple example -- a random walk on
a {\em flat} $d$-dimensional manifold with Cartesian coordinates $\{x^\mu,
\mu=1\ldots d\}$.
At each step the velocity $\vec v$ of the moving `particle'
is random, with an isotropic distribution so that
$\langle v^\mu\rangle =0$ and $\langle v^\mu v^\nu\rangle=
(1/d)\langle{\vec v}\,^2\rangle \delta^{\mu\nu}$.
Hence
\begin{equation}
P(X+\Delta X)=
P(X) + {1\over 2d} (\Delta X)^2 \langle{\vec v}\,^2\rangle\,
\hat\Delta\,P(X) +\ldots
\quad.
\end{equation}
Upon identifying a fictitious `time' $t:=(\Delta X)^2$ and
the diffusion constant $D:=(1/2d)\langle{\vec v}\,^2\rangle$,
we arrive at the diffusion equation $\partial P/\partial t=
D\hat\Delta P$.
This is easily generalized to the (curved) manifold of $S$-matrices.
There, too, increments $\Delta X$ are related to a
fictitious `time' $t$ by
\begin{equation}
t=(\Delta X)^2
\quad.
\label{time_def}
\end{equation}
(Strictly speaking, this relationship holds only
for infinitesimal parameter values;
the exact relationship
between $t$ and finite values of the physical parameter
remains unknown.)
The diffusion equation then reads
\begin{equation}
{\partial\over\partial t}P=D_\beta\hat\Delta_\beta P
\quad.
\end{equation}
Here $\hat\Delta_\beta$ is the Laplace-Beltrami operator
(the generalization of the Laplace operator on curved Riemannian
manifolds), and $D_\beta$ is
the diffusion constant
\begin{equation}
D_\beta={1\over 2d_\beta}\left\langle
 g\left({{\rm d}S\over{\rm d}X}, {{\rm d}S\over{\rm d}X}\right)
\right\rangle
\quad.
\end{equation}
Both are labelled by the symmetry index $\beta$.
The average $\langle g(\ldots)\rangle$ is a measure for
the typical step size of the random walk; it in turn defines a
characteristic `time' scale
\begin{equation}
t_0:= \left\langle
 g\left({{\rm d}S\over{\rm d}X}, {{\rm d}S\over{\rm d}X}\right)
\right\rangle^{-1}
\quad.
\end{equation}
In order to make use of this model, one needs to know
(i) the explicit form of the Laplace-Beltrami operator and
(ii) how diffusion affects the conductance.

The Laplace-Beltrami operator on an arbitrary Riemannian
manifold with coordinates $\{x^\mu\}$ is given by
\cite{barut}
\begin{equation}
\hat\Delta=\sum_{\mu\nu}
{1\over\sqrt{|\det g|}}\,{\partial\over\partial x^\mu}
g^{\mu\nu}\sqrt{|\det g|}{\partial\over\partial x^\nu}
\quad.
\label{laplace}
\end{equation}
Its explicit form on the manifold of
$S$-matrices, with parametrization $\{\tau_a, v^{(i)}\}$,
follows directly from the properties of the metric.
As the metric tensor is composed of a radial and an angular block,
the Laplace-Beltrami operator can be split into two parts,
\begin{equation}
\hat\Delta_\beta = \hat\Delta_{\beta,\tau} + \hat\Delta_{\beta,v}
\quad,
\end{equation}
the first containing derivatives with respect to $\tau$ only,
and the latter containing derivatives with respect to angles only.
The two parts are called radial and angular, respectively.
When applied to the conductance, only the radial part
$\hat\Delta_{\beta,\tau}$ contributes.
This radial part is obtained by inserting (\ref{radial_metric})
and (\ref{volume}) into (\ref{laplace}):
all $\tau$-independent factors cancel,
leaving
\begin{equation}
\hat\Delta_{\beta,\tau} = 2\sum_a {1\over J_\beta}\,
{\partial\over\partial\tau_a}\left[
\tau_a(1-\tau_a)J_\beta {\partial\over\partial\tau_a}\right]
\quad.
\end{equation}
The Laplace-Beltrami operator has two important properties.
First, it is Hermitian in the sense
\begin{equation}
\int{\rm d}\mu(S)\,f\hat\Delta_\beta h =
\int{\rm d}\mu(S)\, h\hat\Delta_\beta f
\quad,
\end{equation}
provided $f$ and $h$ are sufficiently smooth.
Second, the deviation of the dimensionless conductance from its mean,
$\delta_\beta T:=\sum_b\tau_b -\langle T\rangle_\beta$, is an eigenfunction:
\begin{eqnarray}
\hat\Delta_\beta\,\delta_\beta T &=&
\hat\Delta_{\beta,\tau}\,\delta_\beta T
\nonumber \\
&=& 2\sum_{ab}{1\over J_\beta}\,{\partial\over\partial\tau_b}
\left[\tau_b(1-\tau_b)J_\beta{\partial\over\partial\tau_b}
\left(\tau_a-{\beta N^2\over d_\beta}\right)\right]
\nonumber \\
&=& 2\sum_a(1-2\tau_a) + 2\sum_a \tau_a(1-\tau_a)
{\partial\ln J_\beta\over\partial\tau_a}
\nonumber \\
&=& -(d_\beta/N)\,\delta_\beta T
\quad.
\end{eqnarray}

Since the Laplace-Beltrami operator is Hermitian, we are free
to describe the evolution of expectation values
$\langle f\rangle =\int{\rm d}\mu\, Pf$ either in the
`Schr\"odinger picture,' $P\to P(t)$, or
in the `Heisenberg picture,' $f\to f(t)$.
Choosing the latter, we see that
diffusion causes the conductance to approach
its equilibrium value exponentially:
\begin{eqnarray}
\delta_\beta G(t) &:=& \exp[tD_\beta \hat\Delta_\beta]\,\delta_\beta G
\nonumber \\
&=& \exp[-t/(2Nt_0)]\,\delta_\beta G
\quad.
\end{eqnarray}
This immediately yields the conductance-conductance
correlation function
\begin{eqnarray}
C(t) &=& \langle\delta_\beta G\,\delta_\beta G(t)\rangle_\beta
\nonumber \\
&=& C(0)\cdot\exp[-t/(2Nt_0)]
\quad.
\label{c}
\end{eqnarray}
The shape of the weak localization peak follows from a similar
consideration.
At $t=B=0$ the probability density is non-zero only on the
submanifold of time-reversal invariant $S$-matrices:
$P(0)\propto\delta(S^T-S)$.
As $B$ and hence $t$ increase, this distribution diffuses
over the entire manifold of $S$-matrices.
We thus find
\begin{eqnarray}
W(t) &=& \int{\rm d}\mu(S)\,P(t)\,\delta_2 G
\nonumber \\
&=& \int{\rm d}\mu(S)\,P(0)\,\delta_2 G(t)
\nonumber \\
&=& W(0)\cdot\exp[-t/(2Nt_0)]
\quad.
\label{w}
\end{eqnarray}
Equations (\ref{c}) and (\ref{w}) together imply the key result of
this paper:
\begin{equation}
C(t)/C(0)=W(t)/W(0)=\exp[-t/(2Nt_0)]
\quad.
\end{equation}
\section{DISCUSSION}
We are led to the following three conclusions.
(i) The conductance-conductance correlation function in $B$ has the
same shape and width as the weak localization peak:
\begin{equation}
C(B)/C(0)=W(B)/W(0)\quad\forall\,B
\quad.
\end{equation}
This is a stronger version
of the approximate equality (\ref{approx_equal}).
{\em Which} shape the functions have, depends on the relationship between $t$
and $B$.
For finite values of $B$,
this relationship is not known.
(ii) For small $B$, however, we know
from equation (\ref{time_def}) that $t\sim B^2$; hence
\begin{equation}
C(B)/C(0)=
W(B)/W(0)\simeq 1-{\cal O}({B^2})
\quad.
\end{equation}
This is not a very strong result, but it at least
excludes linear line shapes of the kind which are observed
in non-chaotic cavities.\cite{chang}
(iii) Provided we are in the regime where $t\sim B^2$,
the width of both functions increases as $\sqrt{N}$.\cite{assumption}

The first conclusion is consistent with semiclassics only in the limit
$B\to 0$ or $N\to\infty$ (which implies $\alpha_{\rm cl}\to\infty$);
the second conclusion is trivially correct;
and the third conclusion agrees with Hamiltonian
random matrix theory \cite{pluhar} except for very small $N$.
It therefore seems that our Brownian motion model
is exact in the limit of small parameter values
and a large number of channels.
Outside this limit, however, the model can
describe the phenomena at hand only qualitatively, not quantitatively.

The partial failure of the Brownian motion model reveals that
the `walk' on the manifold of $S$-matrices, induced by varying an
external parameter such as $B$, is not really `random':
it cannot be characterized solely by its typical step size.
In other words, the second moment $\langle g({\rm d}S/{\rm d}X,
{\rm d}S/{\rm d}X)\rangle$ is either not appropriate or not sufficient,
or both, to characterize parametric motion;
there must be other constraints.\cite{guess}
To find the nature and physical origin of these constraints, should
be the goal of future investigations.
\acknowledgements
I thank G. Hackenbroich, J. M\"uller, F. von Oppen,
T. Papenbrock, H. A. Weidenm\"uller
and J. A. Zuk for critical reading of the manuscript and helpful
suggestions.
Financial support by the Heidelberger Akademie der Wissenschaften
is gratefully acknowledged.


\begin{references}
\bibitem{reviews}
For reviews see:
S. Washburn and R. A. Webb,
{Adv. Phys.} {\bf 35}, 375 (1986);
C. W. J. Beenakker and H. van Houten in
{\em Solid State Physics}, Vol. 44,
ed. by H. Ehrenreich and D. Turnbull
(Academic Press, New York, 1991);
B. L. Altshuler, P. A. Lee, and R. A. Webb, eds.,
{\em Mesoscopic Phenomena in Solids}
(North-Holland, New York, 1991).
\bibitem{marcus}
C. M. Marcus, A. J. Rimberg, R. M. Westervelt, P. F. Hopkins,
and A. C. Gossard,
{Phys. Rev. Lett.} {\bf 69}, 506 (1992);
M. W. Keller, O. Millo, A. Mittal, and D. E. Prober,
Surf. Sci. {\bf 305}, 501 (1994).
\bibitem{micro}
B. L. Al'tshuler and B. I. Shklovskii,
Zh. Eksp. Teor. Fiz. {\bf 91}, 220 (1986)
$[$Sov. Phys. JETP {\bf 64}, 127 (1986)$]$;
P. A. Lee, A. D. Stone, and H. Fukuyama,
Phys. Rev. B {\bf 35}, 1039 (1987).
\bibitem{transfer}
P. A. Mello,
{Phys. Rev. Lett.} {\bf 60}, 1089 (1988);
C. W. J. Beenakker,
{Phys. Rev. B} {\bf 47}, 15763 (1993).
\bibitem{iwz}
S. Iida, H. A. Weidenm\"uller, and J. A. Zuk,
{Phys. Rev. Lett.} {\bf 64}, 583 (1990);
{Annals of Phys.} {\bf 200}, 219 (1990).
\bibitem{smatrix}
R. A. Jalabert and J.-L. Pichard,
preprint (1994);
R. A. Jalabert, J.-L. Pichard, and C. W. J. Beenakker,
Europhys. Lett. {\bf 27}, 255 (1994);
H. U. Baranger and P. A. Mello,
Phys. Rev. Lett. {\bf 73}, 142 (1994).
\bibitem{jalabert_corr}
R. A. Jalabert, H. U. Baranger, and A. D. Stone,
{Phys. Rev. Lett.} {\bf 65}, 2442 (1990).
\bibitem{baranger_weak}
H. U. Baranger, R. A. Jalabert, and A. D. Stone,
{Phys. Rev. Lett.} {\bf 70}, 3876 (1993).
\bibitem{chang}
A. M. Chang, H. U. Baranger, L. N. Pfeiffer, and K. W. West,
cond-mat/9405077.
\bibitem{hastings}
M. B. Hastings, A. D. Stone, and H. U. Baranger,
cond-mat/9405080.
\bibitem{pluhar}
Z. Pluhar, H. A. Weidenm\"uller, J. A. Zuk, and C. H. Lewenkopf,
preprints (1994).
\bibitem{altland}
A. Altland,
Ph.D. thesis, Universit\"at Heidelberg (1990).
\bibitem{simons}
B. D. Simons, P. A. Lee, B. L. Altshuler,
{Phys. Rev. Lett.} {\bf 70}, 4122 (1993);
E. Br\'ezin and A. Zee,
Phys. Rev. E {\bf 49}, 2588 (1994).
\bibitem{dyson}
F. J. Dyson,
{J. Math. Phys.} {\bf 13}, 90 (1972).
\bibitem{beenakker}
C. W. J. Beenakker,
{Phys. Rev. Lett.} {\bf 70}, 4126 (1993).
\bibitem{equivalence}
A random walk on the manifold of $S$-matrices is certainly not
exactly equivalent to a random walk on the manifold of Hamiltonians.
How the change of base manifolds affects the final results is therefore
not known; however, one would expect equivalence in the
limit of a large number of channels.
\bibitem{landauer}
R. Landauer,
Phil. Mag. {\bf 21}, 863 (1970);
M. B\"uttiker,
Phys. Rev. Lett. {\bf 57}, 1761 (1986).
\bibitem{param}
P. A. Mello and J.-L. Pichard,
J. Phys. I {\bf 1}, 493 (1991);
P. A. Mello and A. D. Stone,
Phys. Rev. B {\bf 44}, 3559 (1991).
\bibitem{symplectic}
The `symplectic case,' $\beta=4$, will not be considered.
\bibitem{coupling}
This assumption is reasonable as long as
(i) the cavity is chaotic;
(ii) dephasing effects (Coulomb interaction, phonons) may be neglected;
and (iii) there are no barriers between cavity and leads, i.e.,
their coupling is maximal.
\bibitem{barut}
A. O. Barut and R. Raczka,
{\em Theory of Group Representations and Applications}
(World Scientific, Singapore, 1986).
\bibitem{assumption}
It is assumed that $t_0$ is independent of $N$.
This is equivalent to assuming $|{\rm d}S_{ij}/{\rm d}X|^2\simeq
{\cal O}(1/N^2)$ for all $i,j$;
or, in the Hamiltonian model \cite{pluhar} $H=H^{(0)}+i\sqrt{t/N} H^{(A)}$,
to assuming $|H_{ij}^{(A)}|^2\simeq {\cal O}(1/N)$.
\bibitem{guess}
That some additional information about the perturbation is needed,
could have been guessed already from the different functional dependences
of $C(\Delta k)$ (Lorentzian; eq. (\ref{mom_corr})) versus
$C(\Delta B)$ (Lorentzian-squared; eq. (\ref{mag_corr})) on the
respective parameter.
%
\end{references}
\end{document}